# Nonvolatile memory with molecule-engineered tunneling barriers

*Tuo-Hung Hou, Hassan Raza, Kamran Afshari, Daniel J. Ruebusch, and Edwin C. Kan*

School of Electrical and Computer Engineering, Cornell University, Ithaca, NY 14853

Email: *th273@cornell.edu*   Tel: 607-255-4181   Fax: 607-254-3508

**ABSTRACT**

We report a novel field-sensitive tunneling barrier by embedding $C_{60}$ in $SiO_2$ for nonvolatile memory applications. $C_{60}$ is a better choice than ultra-small nanocrystals due to its monodispersion. Moreover, $C_{60}$ provides accessible energy levels to prompt resonant tunneling through $SiO_2$ at high fields. However, this process is quenched at low fields due to HOMO-LUMO gap and large charging energy of $C_{60}$. Furthermore, we demonstrate an improvement of more than an order of magnitude in retention to program/erase time ratio for a metal nanocrystal memory. This shows promise of engineering tunnel dielectrics by integrating molecules in the future hybrid molecular-silicon electronics.

In the present charge-based nonvolatile Flash memory technology, the ratio between retention time $t_R$ and program/erase (P/E) time $t_{PE}$ is about $10^{12}$-$10^{14}$. In order to realize this tremendous ratio, field-asymmetric tunneling processes in the tunneling barrier have to be deliberately engineered between retention and P/E. The asymmetry in conventional Flash cells is most often provided by the external P/E voltage. For example, in the NAND Flash, the asymmetry between the Fowler-Nordheim tunneling under P/E and the direct tunneling during retention is exploited. However, this approach limits the scalability of the P/E voltage, which is quickly becoming the major scaling roadblock, considering power dissipation, cycling endurance, and peripheral circuitry design [1-2].

In order to address these issues, the metal nanocrystal (NC) memory has been proposed [3]. Metal NCs enhance the tunneling asymmetry due to the additional band offset between the material-dependent floating-gate work function and Si band edges of the channel. Along with the 3D electrostatic advantages [4], extremely low P/E voltages can be realized [5-6]. Furthermore, tailoring the band structure of the tunneling barrier is another effective way to achieve significant tunneling asymmetry. Various theoretical and experimental approaches based on crested tunneling barriers [7-8], asymmetric layered barriers [5, 9, 10], bandgap-engineered Oxide-Nitride-Oxide (ONO) [11], and double tunnel junction [12-13] have been undertaken. Among them, the double tunnel junction proposed by R. Ohba *et. al.* [12-13] is of particular interest with its superior $t_R$ / $t_{PE}$ ratio at low P/E voltage and the demonstration of excellent memory scalability. This structure consists of a layer of about 1-nm Si NCs sandwiched between two $SiO_2$ layers. These infinitesimal Si NCs are crystallized by annealing a $SiO_2$ / a-Si / $SiO_2$ structure. The size of NCs plays an important role in the memory performance [12]. But its precise control determined by the thin Si layer thickness and total thermal history remains complicated. Therefore, device variation within large memory array and reproducibility from run to run are potential issues.

Molecules with versatile and tunable properties may find many applications in integration with traditional Si technology. We have previously reported the redox states of $C_{60}$ molecules for multi-level charge storage [14]. In this paper, we present a simpler implementation of the double tunnel junction by utilizing the monodisperse nature of these nanoscale entities. $C_{60}$ molecules instead of Si NCs are embedded inside the oxide barrier to overcome the aforementioned limitation on the NC size control. To our best knowledge, it is the first demonstration of molecule-engineered tunneling barrier in Si devices. We will further show improved $t_R / t_{PE}$ ratio in a metal NC memory integrated with this barrier.

The metal-oxide-semiconductor (MOS) capacitors with conventional local oxidation of Si (LOCOS) isolation on p-type substrates were fabricated. After 2.5-nm dry thermal oxidation, $C_{60}$ molecules were thermally evaporated to a thickness of 0.4 to 0.6 nm as measured by the quartz crystal monitor, followed by $SiO_2$ evaporation of 3 nm to complete the tunneling barrier formation. The $C_{60}$ molecules used in this study were obtained commercially (MER Corporation 99.9 %). The area density of $C_{60}$ molecules estimated from the electrical measurement [14] is around $2\times10^{12}/cm^2$. As for the metal NC memory cells, after the tunneling oxide formation, spherical Au NCs were self-assembled on the oxide by the electron-beam evaporation of 1.2-nm Au without annealing. $SiO_2$ control oxide was deposited by plasma-enhanced chemical vapor deposition (PECVD) to a thickness of 30 nm. Finally, a top Cr gate was patterned, followed by $400°C$ forming gas annealing for 30 minutes. The schematics of various heterogonous gate stacks (S1-S5) investigated in this study are illustrated in Fig. 1 (a).

We first examine the gate current through the proposed $C_{60}$-embedded tunneling barrier in Fig. 2. S1 consists of tunneling oxide (2.5-nm thermal $SiO_2$ + $C_{60}$ + 3-nm evaporated $SiO_2$) but not top layers of Au NC and PECVD $SiO_2$. A control sample S2 without the $C_{60}$ layer is also shown for comparison. The field-dependent fit (not shown) confirms that the current transport in S2 is governed by the Frenkel-Poole emission through the shallow traps inside the evaporated $SiO_2$. S1 shows exponential gate current increase by four orders of magnitude due to resonant tunneling through the molecular levels of $C_{60}$. $C_{60}$ molecules are closer to the channel and have a larger and more controllable density than the evaporated $SiO_2$ traps. The gate current saturation above ±3V is limited by high substrate resistance and insufficient minority carrier generation under inversion. Because these field conditions are very far away from those in normal memory operations, the present results are satisfactory for the discussion in this paper. Theoretical tunneling current calculation by the Wentzel-Kramer-Brillouin (WKB) approximation [15] is provided to compare with the experimental data in Fig. 2. The gate current from S1 agrees very well with that from an ideal 2.7-nm $SiO_2$ barrier. It is only 10 times smaller than the calculated current for a single layer of 2.5-nm $SiO_2$, despite much thicker physical thickness provided by the top $C_{60}$ and evaporated $SiO_2$ layers.

The energy band/level diagram of the $C_{60}$-embedded barrier under high-bias conditions, such as the program operation, is illustrated in Fig. 1 (b). The HOMO-LUMO gap (highest occupied molecular orbital , lowest unoccupied molecular orbital) of $C_{60}$ is about 1.64 eV [16] with HOMO and LUMO levels being five-fold and three-fold degenerate, respectively. Furthermore, the specific energy level alignment with the bands of the surrounding dielectrics is determined by interface dipole formation and redox states of $C_{60}$ at thermal equilibrium [14]. Under sufficient external bias, resonant tunneling through $C_{60}$ energy levels is enabled due to the energy of injected electrons from the Si channel exceeding the $C_{60}$ energy levels and Coulomb charging energy. Although a detailed model using the Coulomb blockade theory of single electron tunneling [17] is more complete, a simple two-step tunneling process [18] is sufficiently intuitive to describe the observed phenomena to the first order. The two-step tunneling current density $J$ in the weak coupling regime with accessible energy levels provided by the intermediate $C_{60}$ molecules can be expressed as:

$$J = \sum_{E=E_a}^{\infty} q\, C(E)\, N_t\, \sigma_t\, \frac{P_L(E)\, P_R(E)}{P_L(E) + P_R(E)} \qquad (1)$$

where $N_t$ is the density of $C_{60}$, $\sigma_t$ is the effective capture cross section of $C_{60}$, $P_L$ and $P_R$ are the tunneling probabilities through the left and right oxide barriers, $C$ is the electron source function of the channel. The summation takes into account all electrons with energy higher than the first assessable energy-level of $C_{60}$ molecules $E_a$. Here we assume that the occupancy factors are 1 and 0 for the conduction-band electron states in the channel and gate, respectively. In Fig. 2, the similarity of *J-V* shapes between the 2.5-nm $SiO_2$ and the $C_{60}$-embedded barrier implies *J* is mainly controlled by the left barrier, i.e. $P_L \ll P_R$. This is not surprising because bulk traps in the evaporated $SiO_2$ could greatly enhance $P_R$ through the trap-assisted tunneling process [19]. Very high $C_{60}$ density with a reasonable $\sigma_t = 5 \times 10^{-14}$ $cm^2$ can account for the 10 times current reduction in comparison with the single layer of 2.5-nm $SiO_2$. On the contrary, under low-bias conditions, the resonant tunneling is forbidden due to both the $C_{60}$ HOMO-LUMO gap and the Coulomb charging energy as shown in Fig. 1 (c). The direct tunneling current can be extremely low for a thick barrier and will only be evaluated through the retention measurement in the memory cell discussed below. In reality, the trap-assisted tunneling through the interface states between $C_{60}$ and $SiO_2$ and the bulk traps in evaporated $SiO_2$ may lead to higher current. It is worthwhile mentioning that design optimization for a maximum tunneling asymmetry may be possible by engineering the HOMO-LUMO gap and charging energy of different molecules as well as the dielectric thickness.

The high frequency capacitance-voltage (CV) sweeps with increasing range from $\pm$ 2V to $\pm$ 6V are shown in Fig. 3 for memory cells without the Au NC layer (S3) and with the Au NC layer (S4). Both S3 and S4 are with the $C_{60}$-embedded tunneling barrier. In a separate control sample without both $C_{60}$ and Au NC but with all other dielectric layers, no hysteresis is observed under the same sweep range. In S3, larger negative flatband shifts ($\Delta V_{FB}$) agree with the previous experiment, indicating the preferable hole storage at monoanion $C_{60}^{1-}$ and the higher charge neutrality level (CNL) of interface states between $C_{60}$ and $SiO_2$ [14]. In S4, much larger and symmetric $\Delta V_{FB}$ clearly indicates that both electron and hole can indeed be injected into the upper Au NCs through the resonant tunneling modes provided by $C_{60}$.

Finally, the retention and P/E characteristics of a metal NC memory cell with a single layer of 2.5-nm $SiO_2$ (S5) are compared with S4 in Fig. 4. S4 has longer retention due to the lower escape rate of thermally excited electrons and holes in Au NCs through a physically thicker barrier provided by the additional $C_{60}$ and top $SiO_2$ layers. This is more pronounced for electron storage with at least two orders of magnitude improvement in the extrapolated retention time. It is likely due to the suppression of trap-assisted tunneling of electrons with the high CNL at the $C_{60}/SiO_2$ interface. The P/E speed at ± 10V in S4 is only about 10 times slower than that in S5, in close agreement with the tunneling current results in Fig. 2. Even though the P/E voltage has not yet been optimized, which is expected to scale by improving the coupling ratio with either a thinner or a higher-κ control oxide [5], improved $t_R$ / $t_{PE}$ ratio by at least an order is clearly shown with the field-sensitive $C_{60}$-embedded tunneling barrier. Further improvement would be possible by reducing the non-ideal effects by the $C_{60}/SiO_2$ interface states and the bulk traps in the evaporated $SiO_2$.

We have demonstrated a field-sensitive asymmetry in tunneling probability through the molecule-embedded dielectric from J-V characteristics and from improved $t_R$ / $t_{PE}$ ratio of the memory cells. By taking advantages of versatile and tunable molecular properties, the integration of molecules in Si-based devices provides a simple and promising way to tailor tunneling dielectric properties. Although only the metal NC memory is investigated here, the proposed tunneling barrier can be applied for other charge-based memories such as conventional NAND flash, Si NC, and SONOS memories as well.

This work is supported by National Science Foundation (NSF) through the Center of Nanoscale Systems (CNS). The experimental work is performed at the Cornell Nanoscale Facilities (CNF).

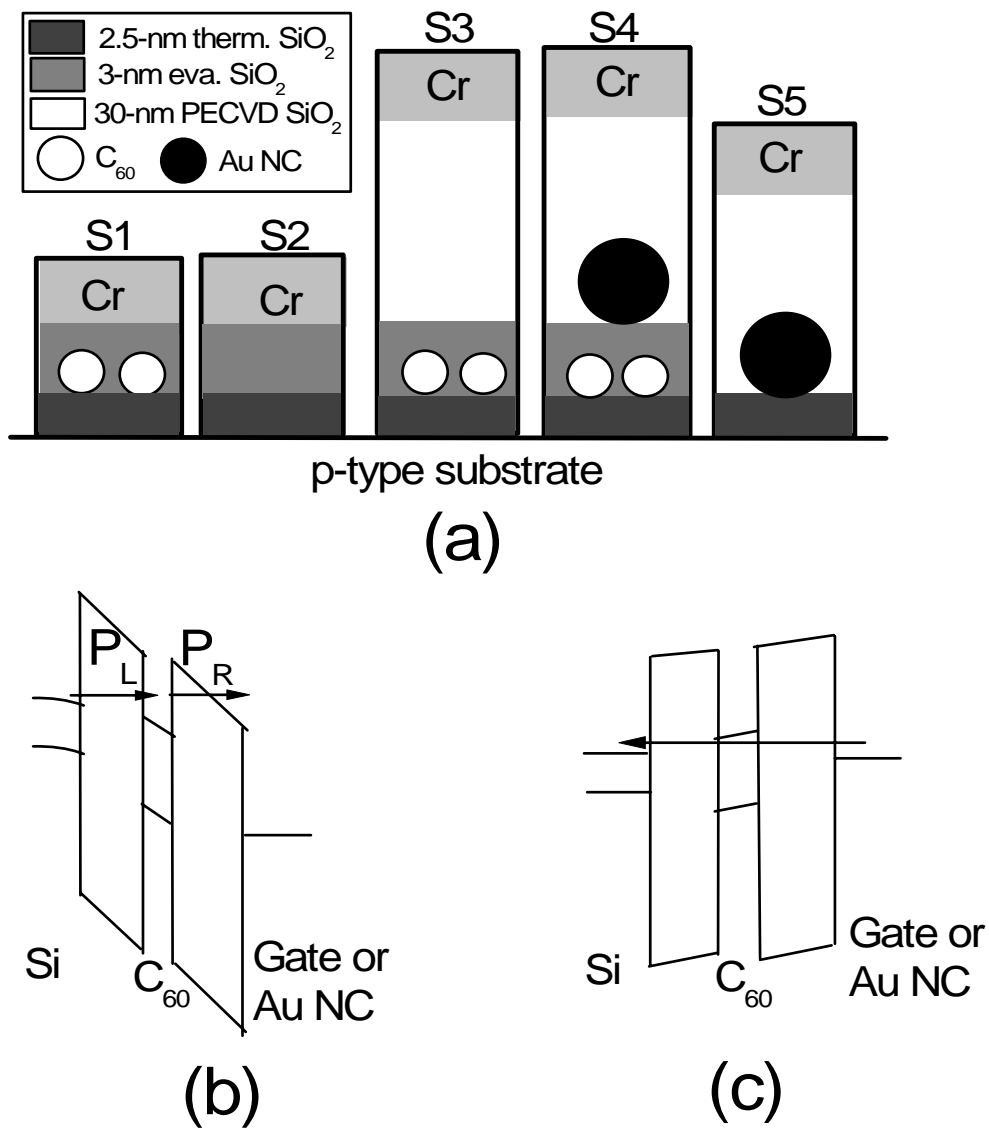

**Figure 1** (a) Schematics of heterogeneous gate stacks (S1-S5) examined in this work. Energy band/level diagram representation of tunneling barriers with (b) resonant tunneling through $C_{60}$ under high electric field, and (c) direct tunneling through $C_{60}$ under low electric field.

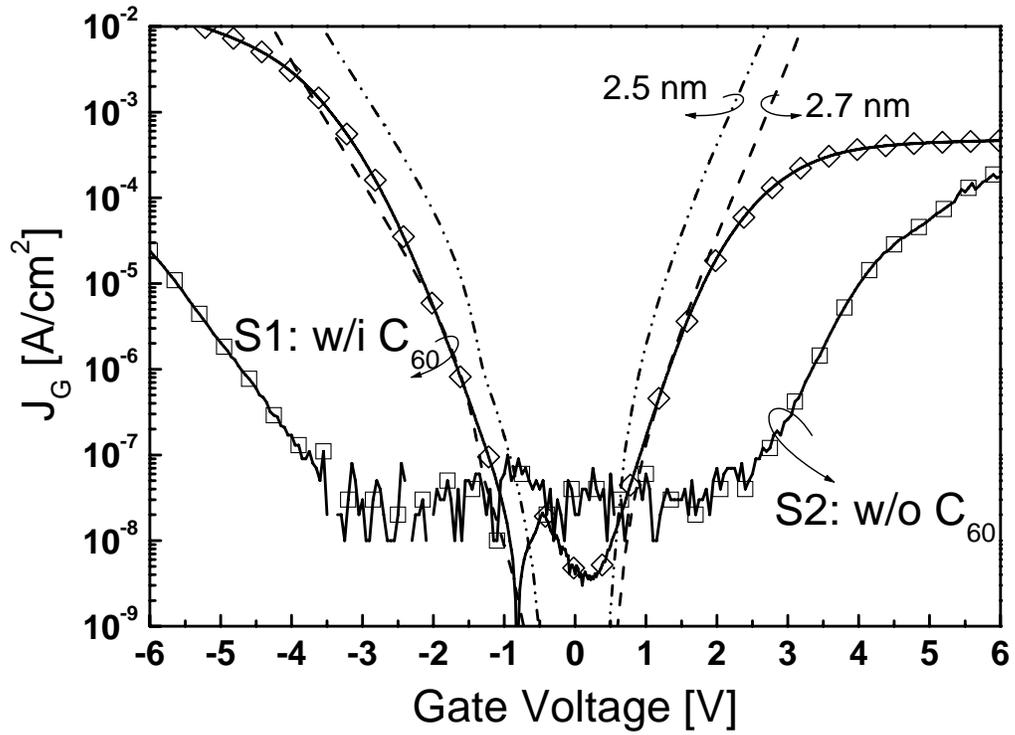

**Figure 2** Measured tunneling current through S1 with a 2.5-nm $SiO_2$ + $C_{60}$ + 3-nm evaporated $SiO_2$ barrier and S2 with a 2.5-nm $SiO_2$ + 3-nm evaporated $SiO_2$ barrier. The dash lines are calculated from the WKB approximation for 2.5-nm and 2.7-nm $SiO_2$.

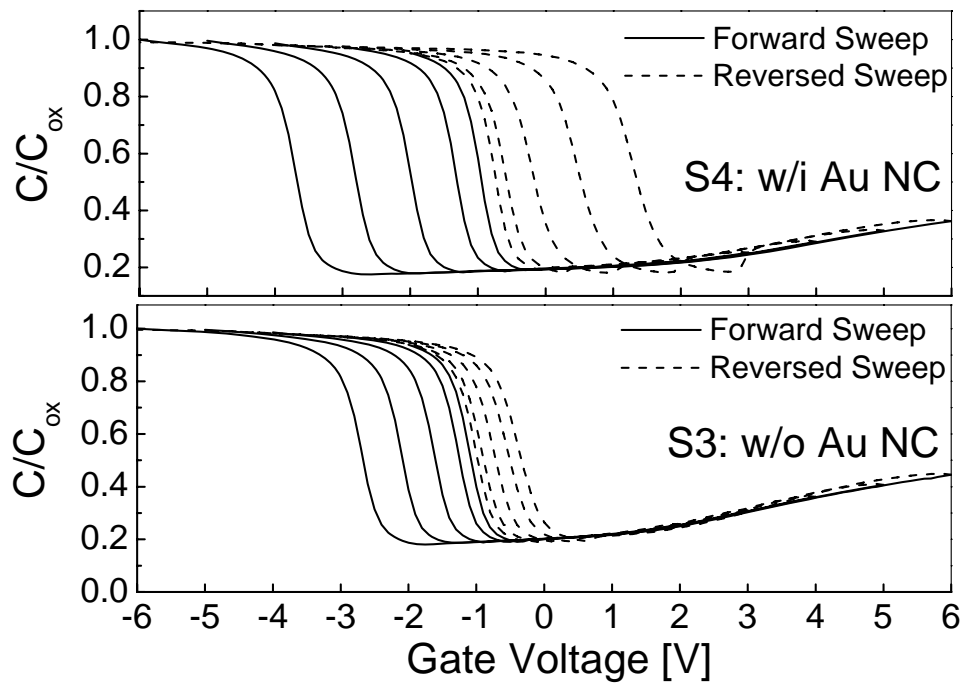

**Figure 3** High frequency CV sweeps with increasing range from ± 2V to ± 6V for memory cells without the Au NC layer (S3) and with the Au NC layer (S4). Both S3 and S4 are with the $C_{60}$-embedded tunneling barrier.

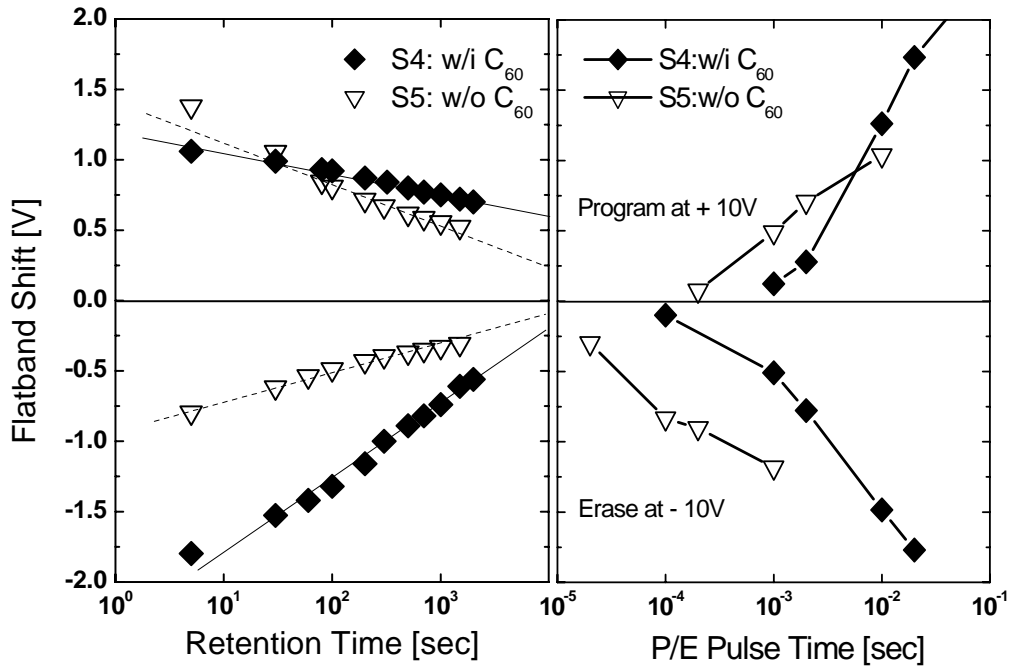

**Figure 4** Retention and P/E characteristics of metal NC memories, S4 with a composite barrier of 2.5-nm $SiO_2$ + $C_{60}$ + 3-nm evaporated $SiO_2$ and S5 with a single layer of 2.5-nm $SiO_2$. For consistent initial conditions, the preset bias prior to the retention and P/E measurements is ± 5V for 3 sec.